\documentclass[letter]{aa}

\usepackage{natbib}
\bibpunct{(}{)}{;}{a}{}{,} 

\usepackage{graphicx}

\usepackage{amssymb,amsmath,bm}
\usepackage{delarray}
\usepackage{mathrsfs}
\usepackage{xcolor}
\usepackage{soul}
\usepackage[varg]{txfonts}
\usepackage{hyperref}
\hypersetup{
  colorlinks   = true, 
  urlcolor     = blue, 
  linkcolor    = blue, 
  citecolor    = blue  
}

\begin{document} 

\title{Kilogauss magnetic fields from simulations of small-scale dynamo action in the convective envelope of a white dwarf}

\authorrunning{Riva et al.}
\titlerunning{Simulations of small-scale dynamo action in a white dwarf model}

\author{Fabio Riva\inst{1,2}\corrauth{fabio.riva@irsol.usi.ch}
       \and
       Tim Cunningham\inst{3}\email{timothy.cunningham@cfa.harvard.edu}
       \and
       Hans-G\"unter Ludwig\inst{4}\email{hludwig@lsw.uni-heidelberg.de}
       \and
       Oskar Steiner\inst{1,5}\email{steiner@leibniz-kis.de}
       \and
       Pier-Emmanuel Tremblay\inst{6}\email{P.Tremblay@warwick.ac.uk}
       }

\institute{
        Istituto ricerche solari Aldo e Cele Daccò (IRSOL), Faculty of Informatics, Università della Svizzera italiana 
        (USI), 6605 Locarno, Switzerland
        \and
        Euler Institute, Universit\`a della Svizzera italiana (USI), 6900 Lugano, Switzerland
        \and
        Center for Astrophysics, Harvard \& Smithsonian, 60 Garden Street, Cambridge, MA 02138, USA
        \and
        Zentrum f\"ur Astronomie der Universit\"at Heidelberg,
        Landessternwarte, K\"onigstuhl 12, D-69117 Heidelberg, Germany
        \and
        Institut f\"ur Sonnenphysik (KIS), D-79104 Freiburg i.~Br., Germany
        \and
        Department of Physics, The University of Warwick, Coventry CV4 7AL, UK
    }

\date{August 06, 2026}
 
\abstract
{
About 20\% of white dwarfs are observed to host large-scale magnetic fields, but the origin of white dwarf magnetism remains uncertain. Small-scale turbulent dynamos (SSDs), which efficiently generate magnetic fields in solar and stellar convection simulations, have so far been studied in white dwarfs only through equipartition arguments in one-dimensional models. We therefore investigate whether turbulent convection in white dwarf surface layers can sustain SSD action through local three-dimensional radiation-magnetohydrodynamics simulations of a DA white dwarf with a convective pure-hydrogen atmosphere, including the full convection zone together with the underlying overshoot and stably stratified layers. Starting from a weak seed field of 1~mG, the magnetic energy undergoes exponential amplification before saturating at a magnetic-to-kinetic energy density ratio of about 5.5\% at the visible surface, demonstrating that SSD action naturally generates kG-strength magnetic fields in convective white dwarf atmospheres. The resulting magnetic field is characterised by a mixed-polarity small-scale structure, with kG field concentrations contributing about 14\% of the total unsigned magnetic flux at the visible surface. Despite a rate of magnetic energy generation amounting to roughly one-seventh of the bolometric flux, no significant modification of the mean stratification is found. Although these fields remain spatially unresolved for observations, they may contribute to spectral line broadening, suggesting that small-scale magnetism in white dwarfs could be more widespread than currently inferred from observations.
}


\AANum{aa62039-26}
\maketitle
\nolinenumbers

\section{Introduction}\label{sec:introduction}
About 20\% of white dwarfs are observed to host large-scale magnetic fields, with strengths ranging from a few kG to several hundred MG~\citep{Bagnulo2021}. The origin of these fields remains uncertain. Strong-field magnetic white dwarfs are, on average, more massive than the general white dwarf population and are often linked to binary interaction or merger scenarios~\citep{Ferrario2020}. Other proposed mechanisms include the survival of magnetic fields from earlier evolutionary phases, such as dynamos operating in giant-star progenitors, and magnetic field generation associated with crystallisation in cooling white dwarfs~\citep{Cantiello2016,Ginzburg2022}. However, no magnetic white dwarf has yet been conclusively linked with a specific field-generation mechanism.

The observed population of magnetic white dwarfs is also affected by detection biases, since weak magnetic fields are difficult to detect, with typical detection thresholds ranging from a few kG to several hundred kG depending on spectral resolution and the availability of spectropolarimetry. Current observations thus mainly constrain organised large-scale fields and do not exclude tangled small-scale magnetic structures at lower field strengths. Magnetism may therefore be more widespread among white dwarfs than is currently inferred from observations.

Small-scale turbulent dynamos (SSDs) are a generic outcome of convection and are thought to operate in many astrophysical systems, including the near-surface layers of solar-type stars~\citep[e.g.][]{Bhatia2022,Riva2024}. In white dwarfs, convection-generated magnetic fields have been estimated from equipartition arguments in one-dimensional models~\citep[e.g.][]{Fontaine1973,Tremblay2015,Yaakovyan2025}, yielding kG strengths but leaving the dynamo saturation level undetermined. Accuracy matters, as white dwarf spectral energy distributions are used for instrumental flux calibration at the 1\% level~\citep[e.g.][]{Elms2024}, where even weak but ubiquitous fields could introduce systematic effects. A self-consistent three-dimensional determination is, however, still lacking.

White dwarfs provide a particularly clean environment for studying SSDs in a self-consistent convective system with minimal interference from artificial lower boundary conditions. This is because their shallow surface convection zones can be fully embedded within the computational domain. In contrast, convection zones in solar and stellar envelope simulations extend deep into the stellar interior and are more strongly affected by artificial boundary conditions within the convection zone.
In this work, we thus perform a radiation-magnetohydrodynamics (R-MHD) simulation of the near-surface layers of a convective pure-hydrogen (DA) white dwarf, showing that SSD action naturally occurs and generates kG magnetic fields.

\section{Numerical setup}
\label{sec:numerical}
We considered a three-dimensional R-MHD simulation representative of the near-surface layers of a pure-hydrogen DA white dwarf with effective temperature $T_\mathrm{eff}\approx12\,000\,\mathrm{K}$ and surface gravity $\log g=8$, computed with the CO5BOLD code~\citep{Freytag2012}. The model was based on a further relaxed version of the $T_\mathrm{eff}\approx12\,000\,\mathrm{K}$ model presented in Table~2 of~\citet{Cunningham2019}, spanning $7.5\times7.5\,\mathrm{km}^2$ horizontally and extending down to approximately $5.6\,\mathrm{km}$ below the visible surface (corresponding to about 7.7 pressure scale heights), thereby encompassing the full convective layer together with part of the underlying stably stratified region.
The initial non-magnetic model was interpolated onto a refined grid of $500\times500\times420$ cells and further relaxed for $3.65\,\mathrm{s}$ of stellar time, after which a uniform vertical magnetic field of strength $1\,\mathrm{mG}$ was introduced as a seed. The simulation was then continued for $19.4\,\mathrm{s}$, corresponding to several convective turnover times (estimated as $\tau_c=H_P/v_\mathrm{z,rms}\approx0.12\,\mathrm{s}$ at the lower Schwarzschild boundary, where $H_P$ is the local pressure scale height and $v_\mathrm{z,rms}$ the root-mean-square vertical velocity), until the magnetic field reached an approximately statistically stationary state. Further details on the numerical setup are given in Appendix~\ref{sec:num}.

We recall that, even in ideal MHD, spatial and temporal discretisation leads to numerical effective viscosity and magnetic diffusivity. 
To reduce the impact of numerical dissipation on the saturation level of the magnetic field, the model was subsequently interpolated onto a higher-resolution grid of $750\times750\times630$ cells, corresponding to a uniform computational cell size of $10\,\mathrm{m}$. After interpolation, a divergence-cleaning step was applied to the magnetic field. The high-resolution simulation was then evolved for an additional $7.9\,\mathrm{s}$. The resulting effective temperature of the model is $T_\mathrm{eff}=12\,032\,\mathrm{K}$, placing it within the ZZ Ceti instability strip.

\section{Small-scale dynamo and magnetic field structure}
\label{sec:results}
The time evolution of the simulation can be summarised as follows. After the introduction of the magnetic seed, the magnetic field grows exponentially over $13.4\,\mathrm{s}$, with a magnetic energy growth rate $\gamma=1.4\,\mathrm{s}^{-1}$, before saturating and reaching a magnetic-to-kinetic energy ratio of $\langle E_\mathrm{m}\rangle_{\tau_\mathrm{R}=1,t}/\langle E_\mathrm{k}\rangle_{\tau_\mathrm{R}=1,t}\approx3.0\%$, where $\langle-\rangle_{\tau_\mathrm{R}=1,t}$ denotes averaging over the surface of constant optical depth $\tau_\mathrm{R}=1$ and time. This is consistent with the behaviour of SSDs in solar and stellar convection simulations~\citep[e.g.][]{Bhatia2022,Riva2024}, characterised by an initial exponential amplification followed by saturation at a fraction of equipartition. We also verified, following the energy transfer analysis by~\citet{Graham2010}, that the mechanism that amplifies magnetic energy is indeed an SSD.

\begin{figure*}[ht!]
    \centering
    \includegraphics[width=0.99\textwidth]{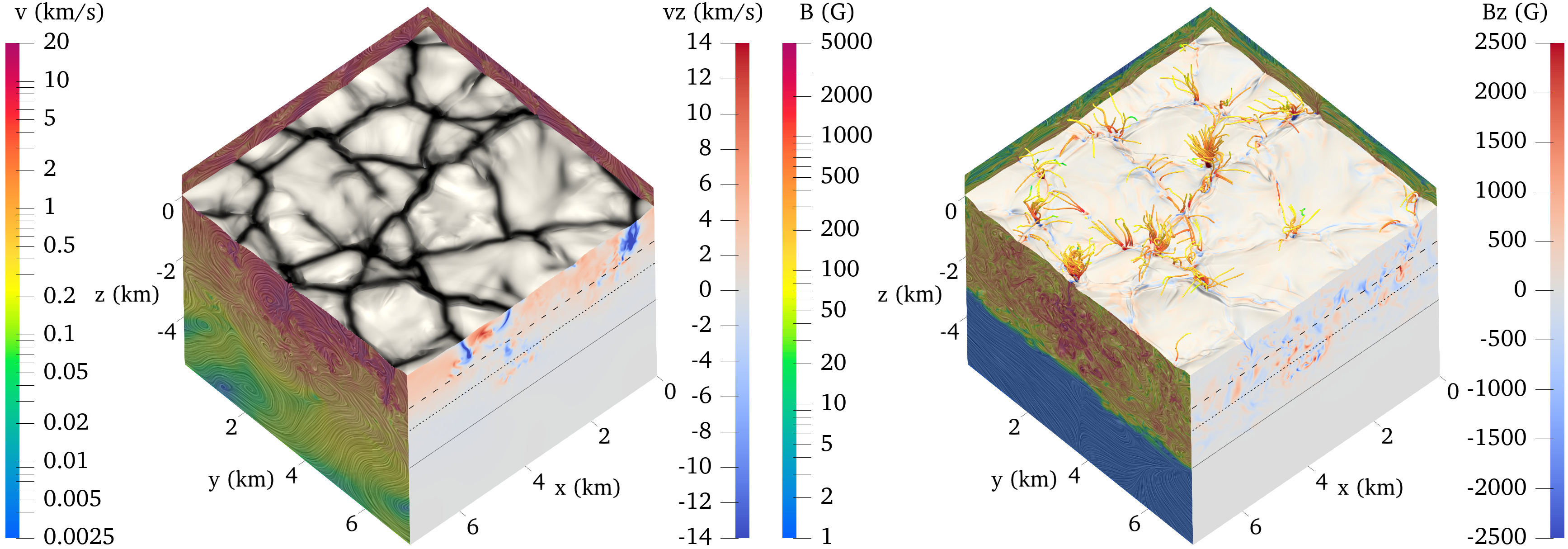}
    \caption{Three-dimensional structure of the velocity and magnetic fields in the saturated state. Left panel: The left and rear sections are colour-coded by the 3D velocity magnitude and overplotted with streamlines, while the front-right section is colour-coded by the vertical velocity. The top grey-scale shows the bolometric intensity on the $\tau_\mathrm{R}=1$ surface. Right panel: The left and rear sections are colour-coded by the magnetic field strength and overplotted with magnetic field lines, while the front-right section and the $\tau_\mathrm{R}=1$ surface are colour-coded by the strength of the vertical magnetic field. Selected magnetic field lines are rendered as tubes above the visible surface and colour-coded by the magnetic field strength. Horizontal black lines on the front-right sections indicate the lower Schwarzschild boundary (dashed), the vanishing enthalpy flux (dotted), and the bottom of the overshoot region~(solid), see text for their definition. Note the large dynamic range represented by the v- and B-colour bars.
    }
    \label{fig:model3D}
\end{figure*}
After interpolation to the higher-resolution grid, the magnetic energy increases further over about $2\,\mathrm{s}$, reaching a new saturation level $\langle E_\mathrm{m}\rangle_{\tau_\mathrm{R}=1,t}/\langle E_\mathrm{k}\rangle_{\tau_\mathrm{R}=1,t}\approx5.5\%$. The analysis presented in this work is based on the saturated phase of the high-resolution run, excluding an initial transient of $2.5\,\mathrm{s}$, and includes a total of 24 three-dimensional snapshots over the final $5.4\,\mathrm{s}$. Following~\citet{Riva2022}, we estimated the Reynolds and magnetic Reynolds numbers in the convection zone, obtaining $\mathrm{Re}\approx1300$ and $\mathrm{Re_m}\approx1500$.

A representative snapshot of the saturated state is shown in Fig.~\ref{fig:model3D}, with the origin of the $z$-axis located at the mean optical depth $\langle\tau_\mathrm{R}\rangle=1$. The velocity field (left panel) displays the familiar granular pattern, with broad upflows below the surface and narrow downdrafts. In contrast, the magnetic field (right panel) exhibits a more intricate, small-scale structure, with magnetic field lines that appear significantly more tangled than the velocity streamlines, consistent with magnetic energy being concentrated at scales smaller than the dominant convective motions, as expected for SSD action.

\subsection{Surface magnetic field}
At the visible surface, the magnetic field is characterised by a mixed-polarity structure with intermittent flux concentrations in the intergranular lanes reaching kG strengths and embedded in a more diffuse background field. Magnetic loops and vortices are also present, as illustrated by the selected magnetic field lines rendered above the surface in Fig.~\ref{fig:model3D}. Quantitatively, the saturated state is characterised by a mean magnetic field strength at the visible surface of $\langle B\,\rangle_{\tau_\mathrm{R}=1,t}=271.7\,\mathrm{G}$.
Peak vertical field strengths reach values of up to $4\,\mathrm{kG}$ in intergranular flux concentrations. These values are noticeably larger than those obtained in SSD simulations of cool main-sequence stars~\citep[e.g.][]{Bhatia2022,Riva2024}, even though the values of $\mathrm{Re}$ and $\mathrm{Re_m}$  attained here are comparable. This likely reflects the strongly compressible nature of white dwarf surface convection, with Mach numbers of about 0.7 at the visible surface, together with the reduced influence of the lower boundary in simulations encompassing the full convection zone.

\begin{figure}
    \centering
    \includegraphics[width=0.95\linewidth]{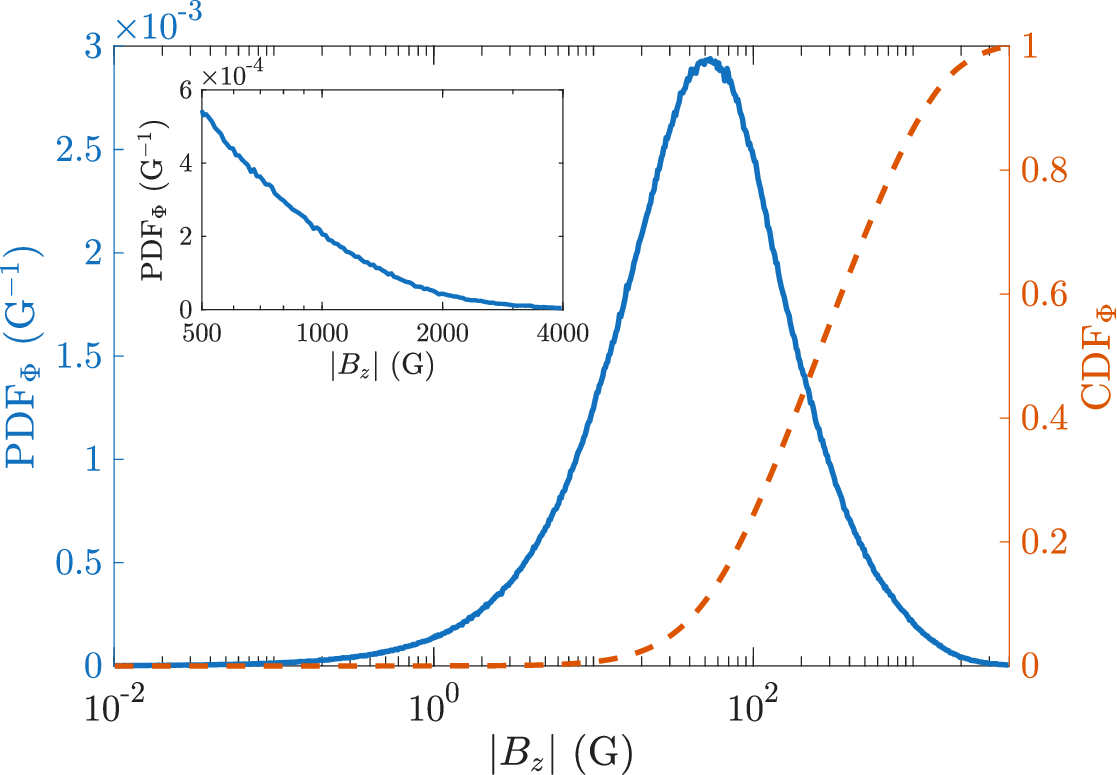}
    \caption{Flux-based probability density function (solid blue line, left axis) and corresponding cumulative distribution function (dashed red line, right axis) as a function of $|B_z|$ at $\tau_\mathrm{R}=1$. The small panel in the top left corner displays a zoom of $\mathrm{PDF}_\Phi$ at large magnetic field strengths.
    }
    \label{Fig:phiBz}
\end{figure}
To quantify how different magnetic field strengths contribute to the unsigned surface magnetic flux, we followed~\citet{Steiner2003} and analysed the flux-weighted probability density function, $\mathrm{PDF}_\Phi(|B_z|)=|B_z|\mathrm{PDF}(|B_z|)/\Phi_\mathrm{tot}$, and the corresponding cumulative distribution function,
$$\mathrm{CDF}_\Phi(|B_z|)=\int_{0}^{|B_z|}\mathrm{PDF}_\Phi(|B_z'|)\mathrm{d}B_z',$$
at $\tau_\mathrm{R}=1$. These are shown in Fig.~\ref{Fig:phiBz}. Here, $\Phi_\mathrm{tot}$ is the total unsigned magnetic flux through the visible surface and $\mathrm{PDF}(|B_z|)$ the probability density function of $|B_z|$. The peak of $\mathrm{PDF}_\Phi(|B_z|)$ is located at $|B_z|\approx53\,\mathrm{G}$, indicating a shift towards magnetic fields two to three times stronger than those obtained by~\citet{Riva2024} for cool main-sequence stars. Moreover, $\mathrm{CDF}_\Phi(|B_z|)$ shows that about half of the unsigned magnetic flux is associated with field strengths $|B_z|<250\ \mathrm{G}$, while kG fields contribute about $14\%$ of the total unsigned magnetic flux.

\subsection{Stratification properties and dynamo action}
To characterise the vertical structure of the simulation, we distinguish three layers, whose locations are indicated in Fig.~\ref{fig:model3D} by horizontal black lines. The lower Schwarzschild boundary, $z_\mathrm{b}$~(dashed lines), marks the bottom of the buoyancy zone and corresponds to the transition from convectively unstable to stable stratification, defined by $\nabla=\nabla_\mathrm{ad}$, with $\nabla=\partial\ln T/\partial\ln P$ 
and $\nabla_\mathrm{ad}$ the adiabatic gradient. The depth where the enthalpy flux vanishes, $z_\mathrm{Fh}$ (dotted lines), separates the convective region from the deeper layers where convective heat transport ceases. The region between $z_\mathrm{Fh}$ and $z_\mathrm{b}$, 
where the stratification is formally stable yet convective motions still carry a net upward enthalpy flux, is known as the Deardorff zone~\citep{Deardorff1966,Kapyla2024}.
Finally, we define a heuristic overshoot depth, $z_\mathrm{o}$~(solid lines), as the depth where the root-mean-square vertical velocity has decreased by $1\,\mathrm{dex}$ relative to its value at $z_\mathrm{Fh}$, providing a purely kinematic boundary to delineate the three dynamical regions relevant for our analysis.\footnote{This differs from the overshoot definition adopted in~\citet{Cunningham2019}, which was based on the diffusion of accreted heavy elements into the stellar interior.}

\begin{figure}
    \centering
    \includegraphics[width=0.95\linewidth]{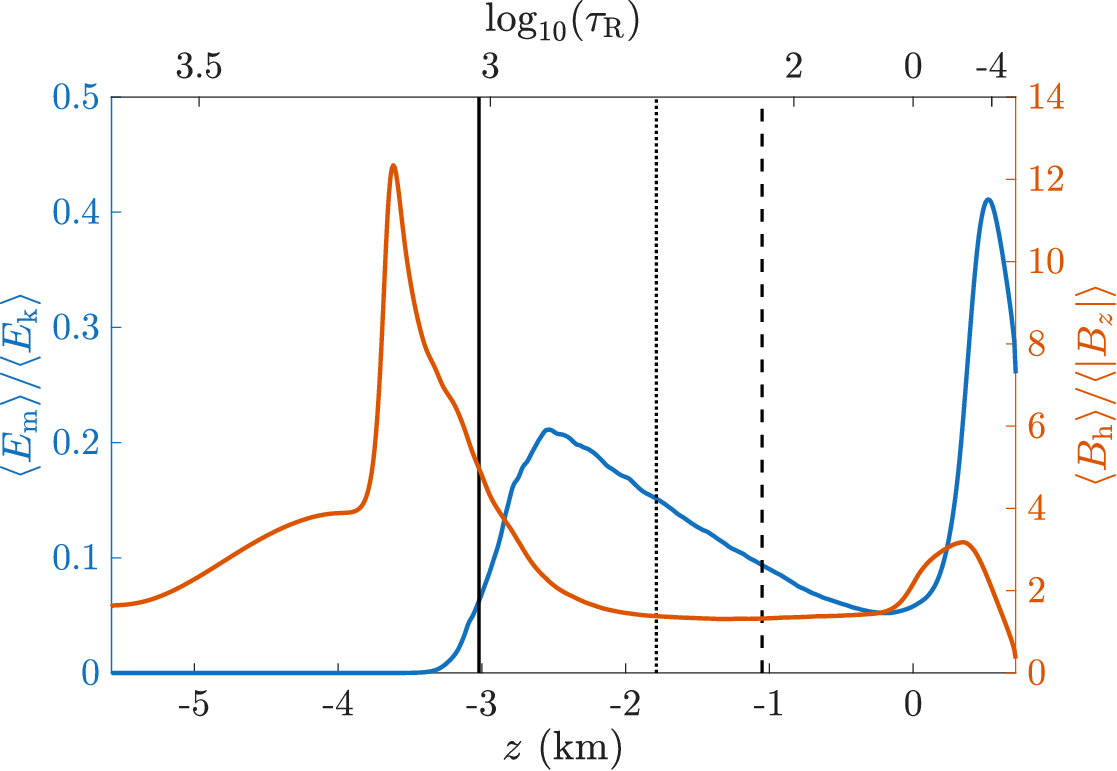}
    \caption{Magnetic field structure as a function of geometrical height. Left axis, blue line: Magnetic-to-kinetic energy density ratio. Right axis, red line: Horizontal to vertical magnetic field strength ratio. Vertical black lines denote the lower Schwarzschild boundary (dashed), the vanishing enthalpy flux (dotted), and the bottom of the overshoot region (solid), see text for their definition. 
    The top horizontal axis indicates selected values of $\log_{10}(\tau_\mathrm{R})$.
    }
    \label{Fig:stratification}
\end{figure}

A qualitative inspection of Fig.~\ref{fig:model3D} already shows that magnetic fields penetrate well below $z_\mathrm{b}$ and are preferentially organised in a horizontal configuration in the deeper layers.
To quantitatively assess how dynamically important magnetic fields are at different depths, and how their geometry changes across the various regions, Fig.~\ref{Fig:stratification} displays the vertical profiles
of $\langle E_\mathrm{m}\rangle/\langle E_\mathrm{k}\rangle$,
and of the anisotropy measure $\langle B_\mathrm{h}\rangle/\langle |B_z|\rangle$, with $B_\mathrm{h}$ the horizontal magnetic field strength (the magnetic field strength, $B$, as a function of optical depth is displayed in Fig.~\ref{Fig:stratification_tauR} in Appendix~\ref{app:app}).
The ratio $\langle E_\mathrm{m}\rangle/\langle E_\mathrm{k}\rangle$, which provides a proxy for the magnetic field strength relative to dynamic equipartition, increases with depth across the convection zone and reaches a local maximum of
$\langle E_\mathrm{m}\rangle/\langle E_\mathrm{k}\rangle=0.21$ 
in the overshoot region, before rapidly decreasing at greater depths.
At the same time, the magnetic field becomes increasingly anisotropic with depth in the overshoot region and below, with $\langle B_\mathrm{h}\rangle/\langle |B_z|\rangle$ rising to values $\gg1$, indicating a strong dominance of horizontal fields in these layers. A similar predominance of horizontal magnetic fields has been reported in solar simulations and observations above the visible surface~\citep[e.g.][]{Steiner2008,Lites2008}, where it is attributed to convective overshoot and the redistribution of magnetic flux by granular flows. In the present model, this behaviour extends both above and below the convection zone, consistent with the transport of magnetic flux by overshooting flows and its subsequent accumulation in stably stratified layers.

To identify where magnetic energy is generated by SSD action, we analysed the work of the Lorentz force
$W_L=\vec{v}\cdot\left[\left(\nabla\times\vec{B}\right)\times\vec{B}\right]/(4\pi)$.
We found that magnetic energy is predominantly amplified in the convection zone, with a maximum close to the lower Schwarzschild boundary, where velocity shear is strongest in the present model. The amplification is primarily associated with turbulent stretching of magnetic fields in downflowing regions, while compression plays only a minor role, consistent with the amplification mechanisms identified in SSD simulations of solar and stellar convection~\citep[e.g.][]{Riva2024}.

Integrating $W_L$ over all depths $z<0$ yields a total rate of magnetic energy generation corresponding to about $14\%$ of the bolometric flux. Despite this non-negligible energy conversion, we do not detect any significant modification of the mean stratification within the statistical uncertainties. 
This is because, in the stationary state, the generated magnetic energy is ultimately converted back into kinetic and thermal energy through the Lorentz force and magnetic dissipation.
The main dynamical effect of the magnetic field is a reduction of the root-mean-square velocity by 2--5\% in the convection zone and up to 10--30\% in the overshoot zone and atmosphere (see Appendix~\ref{app:app}), leading to a corresponding increase in the characteristic advective time-scale.

\section{Discussion and conclusions}
\label{sec:conclusions}
We have presented a three-dimensional R-MHD simulation of the near-surface layers of a DA white dwarf, showing that turbulence in the convection zone efficiently drives SSD action, producing a magnetic-to-kinetic energy density ratio of $5.5\%$ at the visible surface. In the saturated state, about half of the unsigned magnetic flux across the stellar surface is associated with vertical fields stronger than $250\ \mathrm{G}$, with kG fields contributing about $14\%$.
SSD action can therefore generate spatially unresolved kG magnetic fields in the outer layers of convective white dwarfs, suggesting that white dwarf magnetism is more widespread than currently inferred from observations.
However, the present model deliberately targets the regime where the convection zone is shallow enough to be entirely encompassed by the computational domain. From there, it rapidly expands inwards as the star cools~\citep{Bauer2019}, implying that the field strengths found here need not be representative of cooler white dwarfs.
Whether the saturation level relative to equipartition depends on $T_\mathrm{eff}$ remains to be explored.
Moreover, in white dwarfs hosting strong large-scale fossil magnetic fields, SSD-generated fields may be dynamically subdominant or absent.

A key advantage of the present model is that the computational domain encompasses the full convection zone together with the underlying stably stratified layers, thereby minimising the influence of lower boundary conditions. Within this setup, we find that the magnetic field reaches higher strengths at the surface than typically reported in SSD simulations of cool main-sequence stars~\citep[e.g.][]{Bhatia2022,Riva2024}. Magnetic fields also penetrate well below the convection zone and become increasingly dominated by horizontal components in the overshoot region, consistent with the transport of magnetic flux by overshooting flows and its subsequent accumulation as mainly horizontal field in stably stratified layers.
In polluted white dwarfs, the accretion rates inferred from photospheric abundances depend on the efficiency of fingering convection below the convection zone~\citep{Bauer2019}, which magnetic fields can enhance~\citep{Harrington2019,Fraser2024}. An SSD-generated field could therefore bias the inferred accretion rates.

Despite a non-negligible rate of magnetic energy generation, amounting to about $14\%$ of the bolometric flux, we do not detect any significant modification of the mean stratification. This behaviour can be understood in terms of the relatively modest ratio of vertical turbulent pressure to gas pressure in the present model (about 0.1 at the optical surface), which limits the dynamical feedback of magnetic fields on the thermodynamic structure, in contrast to F3V stars where stronger effects have been reported~\citep{Bhatia2022}.

The main limitation of the present study is the finite numerical resolution, which 
limits the attainable Reynolds and magnetic Reynolds numbers. Although the simulation operates above the threshold for dynamo action, the saturation level
of the magnetic field may still depend on the numerical parameters. Higher-resolution simulations will be required to assess the convergence of these quantities and to better constrain the efficiency of magnetic energy transport and dissipation.

The present results also suggest possible observational signatures of SSD-generated magnetic fields in white dwarf spectra. While the predicted fields remain spatially unresolved and are unlikely to produce clearly separated Zeeman components, they may contribute to spectral line broadening.
In the weak-field regime, the additional Zeeman broadening amounts to approximately 0.5\,$\AA$ at H$\alpha$ and 0.3\,$\AA$ at H$\beta$ for the mean surface field of $\langle B\,\rangle_{\tau_\mathrm{R}=1,t}=271.7\,\mathrm{G}$.
Whether this can be disentangled from rotational and turbulent broadening 
will require future dedicated spectral synthesis calculations.
Asteroseismology provides a further constraint on near-surface fields in white dwarfs, with upper limits in some cases below 1 kG~\citep{Rui2025}.
These limits are primarily sensitive to the radial field component, however, whereas the magnetic field found here is increasingly horizontal and decreases rapidly in amplitude below the convection zone~(see Figs.~\ref{Fig:stratification} and \ref{Fig:stratification_tauR}).
SSD-generated fields may therefore remain observationally elusive.

\begin{acknowledgements}
This work was supported by the Swiss National Science Foundation under grant IDs 200020\_182094 and CRSK-2\_237849. The numerical simulations were carried out on Piz Daint at CSCS under project IDs s1172 and u14. P.-E.T. received funding from the European Research Council under the European Union’s Horizon 2020 research and innovation programme number 101002408 (MOS100PC).
\end{acknowledgements}

\bibliographystyle{aa}
\bibliography{bibfile}

\begin{appendix}
\nolinenumbers

\section{Numerical scheme and boundary conditions}
\label{sec:num}
All simulations were computed using the equation of state by~\citet{Tremblay2013} and evolved within the framework of ideal MHD using a Harten–Lax–van Leer (HLL) solver.
The boundary conditions followed those of the reference model of~\citet{Cunningham2019}: periodic in the horizontal directions, with boundaries open at the top to mass flows and radiation and closed at the bottom to mass flux~(i.e. vanishing vertical velocity), while approximately maintaining the prescribed $T_\mathrm{eff}$ through a fixed radiative flux.
Vertical field boundary conditions were imposed at the top and bottom boundaries (i.e. $B_x=B_y=0$ and $\partial_zB_z=0$).

In regions of low plasma$-\beta$ ($\beta<0.1$), the internal energy equation was used instead of the total energy equation to ensure positivity of the gas pressures, at the expense of strict total energy conservation. The Alfvén speed was limited to a maximum of $50\,\mathrm{km}/\mathrm{s}$ to prevent excessively small time steps. Radiative transfer was treated in the grey approximation using mean Rosseland opacities, solved with a long-characteristics Feautrier scheme. More details on the numerical schemes can be found in~\citet{Freytag2012}.

\section{Convective velocity suppression by SSD action} 
\label{app:app}
\begin{figure}[!htbp]
    \centering
    \includegraphics[width=0.95\linewidth]{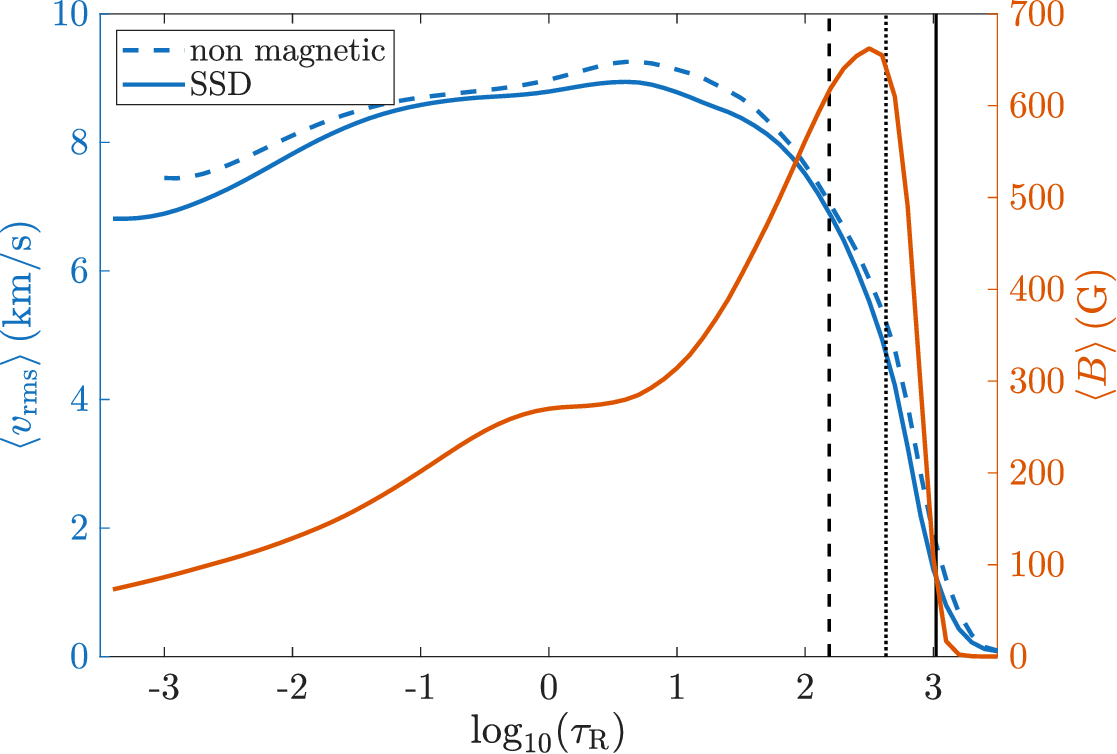}
    \caption{Velocity and magnetic field structure as a function of optical depth. Left axis, blue lines: root-mean-square velocity (all components), $v_\mathrm{rms}$, for both the non-magnetic run (dashed) and the saturated SSD phase (solid). Right axis, red line: Magnetic field strength. Vertical black lines denote the lower Schwarzschild boundary (dashed), the vanishing enthalpy flux (dotted), and the bottom of the overshoot region~(solid), see text for their definition.
    }
    \label{Fig:stratification_tauR}
\end{figure}
Figure~\ref{Fig:stratification_tauR} shows vertical profiles of the root-mean-square velocity, $v_\mathrm{rms}$ (in blue), for both the non-magnetic run~(dashed) and the saturated SSD phase (solid), as well as the mean magnetic field strength $B$~(solid, red), as a function of $\log_{10}(\tau_\mathrm{R})$. The magnetic field acts to suppress convective motions throughout the atmosphere and convection zone. Comparing the non-magnetic (dashed blue) and saturated (solid blue) profiles of $v_\mathrm{rms}$, the reduction amounts to 2--5\% within the convection zone and reaches up to 10--30\% in the overshoot region and atmosphere, consistent with the increasing magnetic-to-kinetic energy ratio found at depth (see Fig.~\ref{Fig:stratification}).\footnote{We note that the different resolution between the reference non-magnetic case and the SSD saturation phase might also affect the amplitude of velocity fluctuations.} The magnetic field strength profile~(solid red, right axis) peaks near the bottom of the convection zone, where SSD amplification is most efficient, before decreasing rapidly in the stably stratified layers below.
The maximum mean field strength, about $660\,\mathrm{G}$, exceeds the surface value by a factor of roughly 2.4. This inward increase is qualitatively consistent with the equipartition estimates of~\citet{Tremblay2015} and \citet{Yaakovyan2025}, which also peak below the photosphere, since the equipartition field grows with depth as the density rises faster than the convective velocity decreases. Quantitatively, however, the dynamo saturates well below equipartition, reaching 5.5\% of the kinetic energy density at the visible surface and up to 21\% in the overshoot region, as discussed in Sect.~\ref{sec:results}.

\end{appendix}

\end{document}